\documentstyle[12pt,epsfig,amsmath,amssymb,psfrag,pifont,cite]{article}
\numberwithin{equation}{section}
\numberwithin{table}{section}
\newcommand{\lambdabf}{{\mbox{\boldmath $\lambda$}}}

\renewcommand{\d}{\partial}

\def\ga{\mathrel{\raise.3ex\hbox{$>$\kern-.75em\lower1ex\hbox{$\sim$}}}}
\def\la{\mathrel{\raise.3ex\hbox{$<$\kern-.75em\lower1ex\hbox{$\sim$}}}}

\def\I_M{{I_{\scriptscriptstyle M\times M}}}

\def\be{\begin{equation}}
\def\ee{\end{equation}}
\def\bea{\begin{eqnarray}}
\def\eea{\end{eqnarray}}

\newcommand{\beqal}{\begin{eqnarray}\label}
\newcommand{\beqa}{\begin{eqnarray}}
\newcommand{\eeqa}{\end{eqnarray}}

\begin{document}

\begin{titlepage}
\begin{center}

\vskip .2in

{\Large \bf \textbf{Universal thermal and electrical conductivity from holography}}
\vskip .5in

{\bf Sachin Jain\footnote{e-mail: sachjain@iopb.res.in}\\
\vskip .1in
{\em Institute of Physics,\\
Bhubaneswar 751~005, India.}}
\end{center}\noindent
\baselineskip 15pt

\begin{center} {\bf ABSTRACT}

\end{center}
\begin{quotation}\noindent
\baselineskip 15pt
It is known from earlier work of Iqbal, Liu \cite{Iqbal:2008by} that the boundary transport coefficients such as electrical conductivity (at vanishing chemical potential), shear viscosity etc. at low frequency and finite temperature can be expressed in terms of geometrical quantities evaluated at the horizon. In the case of electrical conductivity, at zero chemical potential gauge field fluctuation and metric fluctuation decouples, resulting in a trivial flow from horizon to boundary. In the presence of chemical potential, the story becomes complicated due to the fact that gauge field and metric fluctuation can no longer be decoupled. This results in a nontrivial flow from horizon to boundary. Though horizon conductivity can be expressed in terms of geometrical quantities evaluated at the horizon, there exist no such neat result for electrical conductivity at the boundary. In this paper we propose an expression for boundary conductivity expressed in terms of geometrical quantities evaluated at the horizon and thermodynamic quantities. We also consider the theory at finite cutoff recently constructed in  \cite{Bredberg:2010ky}, at radius $r_{c}$ outside the horizon and give an expression for cutoff dependent electrical conductivity $(\sigma(r_c))$, which interpolates smoothly between horizon conductivity $\sigma_{H}(r_c \rightarrow r_h)$ and boundary conductivity $\sigma_{B}(r_c \rightarrow \infty)$. Using the results about the conductivity we gain much insight into the universality of thermal conductivity to viscosity ratio proposed in  \cite{Jain:2009bi}.

\end{quotation}
\vskip 2in
\end{titlepage}
\vfill
\eject

\setcounter{footnote}{1}
\section{Introduction}
AdS/CFT correspondence provides us a powerful tool to study the gauge theories at strong coupling. In particular, we can gain insight into transport properties of gauge theories by studying their gravity duals. At finite temperature and at large length scales, any interacting QFT is described by hydrodynamics. In the gravity side, finite temperature amounts to having a black hole and long wave length physics of field theory is governed by the near horizon physics of the black hole. This idea was employed in  \cite{Iqbal:2008by} to show that, in the low frequency limit, the linear response of the boundary theory is captured completely by the near horizon physics. In \cite{Iqbal:2008by}, authors studied transport coefficients which corresponds to the mass less modes in the bulk, resulting in trivial flow from horizon to boundary. This resulted in an equality between the boundary and the horizon transport coefficients. So when there is a nontrivial flow from horizon to boundary (like massive bulk modes), horizon physics will no longer be able to capture the whole low frequency AdS/CFT response. Calculation of electrical conductivity in the presence of non-zero chemical potential is one such example  where corresponding mode in the bulk shows a non trivial flow from horizon to boundary. These flow equations are in general complicated second order differential equations (if more than one charge is present they are coupled as well) and apriori there is no reason that electrical conductivity for  different theories will show some universal features. The aim of this paper is to investigate whether there exist universal result for the electrical conductivity. It turns out that the conductivities for gauge theories dual to $R-$charge
 black hole in $4, 5$ and $7$ dimension and Reissner-Nordstrom black hole in any dimension behaves in a universal manner.

The paper is structured as follows. Section $2$ is a review of earlier work \cite{Jain:2009pw, Myers:2009ij}. We set it up in way that allows us to generalize some of the earlier results. In section $3$, we take up several examples such as $R-$charge
 black hole in $4, 5$ and $7$ dimensions and Reissner-Nordstrom black hole in any dimension to demonstrate the relation between horizon and boundary conductivity. In section $4$ we concentrate more on Reissner-Nordstrom black hole in arbitrary dimension  and study cutoff dependence of electrical conductivity. Results of this section section can be generalized to much more general systems. In section $5$, we study imaginary part of electrical conductivity, which is shown to be fixed once the real part of electrical conductivity is known.  In section $6$, we study thermal conductivity and thermal conductivity to viscosity ratio. In section $7$, we compute the electrical conductivity, thermal conductivity and thermal conductivity to viscosity ratio of gauge theories dual to charged and uncharged Lifshitz like black holes. The paper ends with discussion of our results. Appendix $A$  contains some discussion on how to relate boundary CFT conductivity to universal conductivity of streached horizon. Section $B$ contains discussion on thermal conductivity. Section $C, D$ contains discussion on thermal conductivity to viscosity ratio.

\section{General perturbation equation}
In this section we write down perturbation equation required for computation of electrical conductivity (for details see \cite{Jain:2009pw}
). For that we consider an action of the form 
\begin{eqnarray}
S &=&  \int d^{d+1}x~~\sqrt{-g}(\frac{1}{2 \kappa^{2}} 
R  - \frac{1}{4 g_{d+1}^{2}} \widehat{G}_{IJ}(r) F_{\mu\nu}^I
 F^{\mu \nu\, J} + \rm{Scalar~ field~ terms +....})
\nonumber\\
&=& \frac{1}{2 \kappa^{2}}\int d^{d+1}x~~\sqrt{-g}(
R  - \frac{1}{4} G_{IJ}(r) F_{\mu\nu}^I
 F^{\mu \nu\, J} + \rm{Other~ terms}),
\label{lagrangian}
\end{eqnarray}
where $ F^{\mu \nu\, J}$ is the field-strength tensor of the $J$-th $ U(1) $ gauge field and $G_{IJ} = \frac{2 \kappa^{2}}{g_{d+1}^{2}}\widehat{G}_{IJ}$.  The metric that we take is of the form
\begin{equation}
 ds^{2} = g_{tt}(r) dt^{2} + g_{rr}(r) dr^{2} + g_{xx}(r) \sum_{i=1}^{d-1} (dx^{i})^{2},\label{metric11}
\end{equation} where $r$ is the radial coordinate. We have assumed full rotational symmetry in $x^{i}$ directions (so that $g_{ij} = g_{xx} \delta_{ij}$, where $i,j$ indices runs over all indices except $r, t$). We also assume that metric components  depends on 
radial coordinate only. We shall work with metric\footnote{For charged black holes, there will exist inner horizons.} which has an event horizon, where $g_{tt}$ has 
a first order zero and $g_{rr}$ has a first order pole where as other metric components are finite (as well as non vanishing) at the horizon. The boundary of the 
space time is at $r = \infty$. The gauge coupling $G_{IJ}$ may be constant or in general can be a function of $r$, where $r$ dependence might come through scalar field.
  
 Maxwell equation can be written as
\begin{equation}
\partial_\mu \Big(\sqrt{- g}G_{IJ}F^{\nu \mu}_{J}\Big) = 0.
\end{equation}
If we consider $G_{IJ}$ to be diagonal and only $A_{t}(r)$ component to be non zero, we can define charge density to be,
\begin{equation}
\rho_{I} = \frac{1}{2 \kappa^{2}} \sqrt{- g}G_{II}g^{rr}g^{tt}F^{I}_{r t} .\label{rho}
\end{equation}
Since our aim is to compute the electrical conductivity using Kubo formula, it is sufficient to consider perturbations in the tensor (metric)  and the vector (gauge fields) modes around the black hole solution and keep other fields such as scalars unperturbed. So perturbations are of the form:
\begin{equation}
g_{\mu\nu}= {\bf g}^{(0)}_{\mu\nu} + h_{\mu\nu}~,
\quad\quad
A^{I}_\mu = {\bf A}^{I(0)}_\mu + {\cal A}^I_\mu~
.
\end{equation}
where ${\bf g}^{(0)}_{\mu\nu}$ and ${\bf A}^{I(0)}_\mu $ are background metric and gauge fields.
In order to determine electrical
 conductivity it is enough to  consider perturbations in $(tx^{1})$ and
$(x^{1}x^{1})$ component of the metric tensor and $x^{1}$ component of the gauge fields. Moreover one can
choose the perturbations to depend on radial coordinate $r$, time $t$ and one of the spatial coordinates say $x^{2}$. 
 A convenient ansatz, with the above restrictions in mind, is
\begin{equation}
h_{tx^{1}} = {\bf g}_{0x^{1}x^{1}}~T(r)~e^{-i\omega t + i q x^{2}},\quad
h_{x^{2}x^{1}} = {\bf g}_{0xx}~Z(r)~e^{-i\omega t + i q x^{2}},\quad
{\cal A}^I_{x^{1}} =  \phi_{I}(r)~~e^{-i\omega t + i q x^{2}}.\quad
\end{equation}
Here $\omega$ and $q$ represent the frequency and the momentum in $x^{2}$ direction respectively and we set perturbations in the other components to be equal to zero. Next step is to find linearized equations which follow from the equations of motion. It turns out that at the level of linearized equation and at zero momentum limit metric perturbation $Z(r)$ decouple from the rest. One can further eliminate $T(r)$ reducing it to a equation for perturbations in gauge fields only. After substitution one finds the equations for perturbed gauge fields to be
\begin{equation}
\frac{d}{dr}(N_I\frac{d}{dr}\phi_I(r))-\omega^2 N_I~ g_{rr} g^{tt} \phi_I(r)+\sum\limits_{J=1}^m  M_{IJ}\phi_J(r)=0.
\label{eqnmotion}
\end{equation}
with
\begin{equation}
N_I=\sqrt{-g}G_{II}g^{xx}g^{rr}.\label{N}
\end{equation}
and
\begin{equation}
 M_{IJ}=F_{rt}^I \sqrt{-g}G_{II}g^{xx}g^{rr}g^{tt}G_{JJ}F_{rt}^J.
\end{equation}
Note that $ M_{IJ}= M_{JI}$.
Let us note that, using eqn. (\ref{rho}), we can write
\begin{equation}
 M_{IJ} = (2\kappa^{2})^{2} \rho_{I} \rho_{J} \frac{g_{rr} g_{tt}}{\sqrt{-g} g_{xx}}.\label{eqnmotion0}
\end{equation}
For evaluating the conductivity in the low frequency limit and for non-extremal backgrounds, we only need to solve equations up to zeroth order in $\omega$. To that order one finds,
 \begin{equation}
\frac{d}{dr}(N_I\frac{d}{dr}\phi_I(r))+\sum\limits_{J=1}^m M_{IJ}\phi_J(r)=0.\label{eqnmotion1}
 \end{equation}
Following \cite{Jain:2009pw}, we now write down the effective action which reproduces the eqn.(\ref{eqnmotion1}) and  extract out the expression for electrical conductivity using Kubo formula.
\subsection{Effective action and expression for conductivity}
The electrical conductivity is usually computed from current-current correlator
\begin{eqnarray}
 \lambdabf &=& - \lim_{\omega \rightarrow 0} \frac{G_{xx}(\omega, q=0)}{i \omega}\nonumber\\
  &=& \lim_{\omega \rightarrow 0}\frac{1}{2 \omega}\int_{- \infty}^{\infty}dt~~e^{-i \omega t}\int d\vec{x}\langle[J_{x}(t,\vec{x}), J_{x}(0,\vec{0})]\rangle .\end{eqnarray} The current-current correlator can be computed by taking second derivative of effective action which reproduces the eqn.(\ref{eqnmotion1}) with respect to boundary fields \cite{Son:2002sd}. The expression for electrical conductivity can formally be written as $\lambda = i \sigma_{0}+ \sigma.$
\begin{itemize}
 \item {\textbf{$\sigma (=\Re(\lambda))$:}} In order to determine the real part of the conductivity ($\sigma$), we follow \cite{Jain:2009pw}.
Effective action can be written as
\begin{eqnarray}
S &=& \frac {1}{2 \kappa^{2}} \int \frac{d^{d}q}{(2\pi)^d}dr \Big[\frac{1}{2}\sum\limits_{I=1}^m N_I(r)\frac{d}{dr}\phi_{I}(r,\omega,q)\frac{d}{dr}\phi_{I}(r,-\omega,-q)\nonumber \\
&+&\frac{1}{2}\sum\limits_{I,J=1}^m M_{IJ}(r)\phi_{I}(r,\omega,q)\phi_{J}(r,-\omega,-q)  \Big].
\end{eqnarray}
Boundary action is given by
\begin{eqnarray}
S_\epsilon &=&\lim_{r\rightarrow \infty}\frac {1}{2 \kappa^{2}} \int \frac{d^{d}q}{(2\pi)^d}\left(\frac{1}{2}\sum\limits_{I=1}^m N_I(r)\frac{d}{dr}\phi_{I}(r,\omega,q)\phi_{I}(r,-\omega,-q)\right)\nonumber\\
&=&\lim_{r\rightarrow \infty}\int \frac{d^{d}q}{(2\pi)^d}\sum\limits_{I\geq K, I,K=1}^m\phi_{I}^{0}(\omega, q){\cal F}_{IK} (\omega,q)\, \phi_{K}^{0} (-\omega, -q)
.\end{eqnarray} where the boundary value of the field $\phi_{I}(r)$ is $\phi^{0}_{I}(\omega, q)$.
Next, the retarded correlators
are given by
\begin{equation}
   G^R   \;=\;
  \begin{cases}  - 2  {\cal F}_{JK} (\omega,q)\,,
                                                    & J=K , \\
           \noalign{\vskip 4pt}
           -  {\cal F}_{JK} (\omega,q)\,,
                                                    & J\neq K . 
         \end{cases}
   \label{f_fth}
\end{equation}
The expression for diagonal and off diagonal parts 
  of the conductivity can be written as 
\begin{eqnarray}
\sigma_{II}&=&-\lim_{\omega\rightarrow 0}\frac{\Im \Bigg(G^R(\omega,q=0)\Bigg)}{\omega}\nonumber\\
&=&\frac {2 ~\Im{ \Bigg({\cal F}_{II} (\omega\rightarrow 0,q=0)\Bigg)}}{\omega},\label{defsig}
\end{eqnarray}
and
\begin{eqnarray}
\sigma_{IJ}&=&-\lim_{\omega\rightarrow 0}\frac{\Im \Bigg(G^R(\omega,q=0)\Bigg)}{\omega}\nonumber\\
&=&\frac { \Im{ \Bigg({\cal F}_{IJ} (\omega\rightarrow 0,q=0)\Bigg)}}{\omega},
\end{eqnarray}

respectively.

In order to find out $\Im\Big( {\cal F}\Big)$, we need to compute,  
\begin{equation}
 \Im\Big[\lim_{r\rightarrow \infty}\frac {1}{2 \kappa^{2}} \int \frac{d^{d}q}{(2\pi)^d}\left(\frac{1}{2}\sum\limits_{I=1}^m N_I(r)\frac{d}{dr}\phi_{I}(r,\omega,q)\phi_{I}(r,-\omega,-q)\right)\Big].
\label{bdyaction}
\end{equation}
Now
\begin{eqnarray}
&&\frac{d}{dr}\Im \Big(\sum_{I=1}^m N_I(r)\frac{d}{dr}\phi_{I}(r,\omega,q)\phi_{I}(r,-\omega,-q)\Big)\nonumber\\
&&=\Im \Big[\sum_{I=1}^m \frac{d}{dr}(N_I(r)\frac{d}{dr}\phi_{I}(r,\omega,q))\phi_{I}(r,-\omega,-q)\nonumber\\
&&+\sum_{I=1}^m N_I(r)\frac{d}{dr}\phi_{I}(r,\omega,q)\frac{d}{dr}\phi_{I}(r,-\omega,-q)\Big].
\end{eqnarray}
Using (\ref{eqnmotion1}), r.h.s of above equation reduces to
\begin{equation}
\Im\Big[-\sum\limits_{I,J=1}^m M_{IJ}(r)\phi_{I}(r,\omega,q)\phi^J(r,-\omega,-q)+\sum\limits_{I=1}^m N_I(u)\frac{d}{dr}\phi_{I}(r,\omega,q)\frac{d}{dr}\phi_{I}(r,-\omega,-q)\Big]  ,
\end{equation}
which is equal to zero since the quantity in the bracket is real.
 Then (\ref{bdyaction}) can as well be calculated at the horizon i.e. at $r = r_{h}$. This simplifies calculations significantly. 
Regularity at the horizon implies
\begin{equation}
\lim_{r\rightarrow r_{h}}\frac{d}{dr}\phi_I(r)=-i \omega \lim_{r\to r_{h}}\sqrt{\frac{g_{rr}}{g_{tt}}} \phi_{I}(r)+{\mathcal{O}}(\omega^{2}). 
\end{equation}
Hence (\ref{bdyaction}) reduces to 
\begin{equation}
\Im\Big[-i \omega\lim_{r\rightarrow r_{h}}\frac {1}{2 \kappa^{2}} \int \frac{d^{d}q}{(2\pi)^d}\sqrt{\frac{g_{rr}}{g_{tt}}}\left(\frac{1}{2}\sum\limits_{I=1}^m N_I(r)\phi_{I}(r,\omega,q)\phi_{I}(r,-\omega,-q)\right)\Big].
\end{equation}

Let us note that, if we take  the solutions  of the form
 \begin{equation}
\phi_{I}(r,\omega,q) = \sum\limits_{A=1}^m \psi^{I}_{A}(r,\omega,q)\phi^{0}_{A} ,
 \end{equation}
  where
 \begin{equation}
 \lim_{r\rightarrow \infty}\phi_{I}(r,\omega,q) = \phi^{0}_{I},
 \end{equation}then we get
 \begin{equation}
 \Im{\Big( {\cal F}_{II}\Big) }= \omega\frac{1}{2 \kappa^{2}}\sqrt{\frac{g_{rr}}{g_{tt}}}\sum\limits_{A=1}^m  \frac{1}{2} N_{A} \psi^{I}_{A}(r)\psi^{I}_{A}(r)\Bigg|_{r_{h}}, 
 \end{equation}and 
\begin{equation}
 \Im{ \Big({\cal F}_{IJ}\Big)} = \omega\frac{1}{2 \kappa^{2}}\sqrt{\frac{g_{rr}}{g_{tt}}}\sum\limits_{A=1}^m  N_{A} \psi^{I}_{A}(r)\psi^{J}_{A}(r)\Bigg|_{r_{h}}.
 \end{equation}
 
\item {\textbf{Single charge case:}} For single charge case, consider $\phi(r) = \psi(r) \phi_{0}$, then we get
 \begin{equation}
 \Im{\Big( {\cal F}\Big) }= \omega\frac{1}{2 \kappa^{2}}\sqrt{\frac{g_{rr}}{g_{tt}}}  \frac{1}{2} N_{1} \psi(r)\psi(r)\Bigg|_{r_{h}}. 
 \end{equation} Using eqn.(\ref{defsig}), we get 
\begin{eqnarray}
\sigma &=& \frac {1}{2\kappa^{2}} \sqrt{\frac{g_{rr}}{g_{tt}}}  N_{1} \psi(r)\psi(r)\Bigg|_{r_{h}} \nonumber\\
 &=& \frac {1}{2 \kappa^{2}} G_{11}(r)~ g_{xx}^\frac{d-3}{2} \psi^{2}(r)\Big{|}_{r=r_{h}}\nonumber\\
&=& \sigma_{H}~~\psi^{2}(r)\Big{|}_{r=r_{h}} ,\label{sincleconduc}
\end{eqnarray}where \begin{eqnarray}
\sigma_{H} &=& \frac {1}{2 \kappa^{2}} G_{11}(r)~ g_{xx}^\frac{d-3}{2}\Big{|}_{r=r_{h}} \nonumber\\
&=& \frac {1}{g_{d+1}^{2}} \widehat{G}_{11}(r)~ g_{xx}^\frac{d-3}{2}\Big{|}_{r=r_{h}}
.\label{horisig}\end{eqnarray} We can also compute  conductivity at any arbitrary radius say at $r_{c}$. This is given by 
\begin{equation}
 \sigma(r_{c}) = \frac {1}{2\kappa^{2}} G_{11}(r)~ g_{xx}^\frac{d-3}{2}\Big{|}_{r=r_{h}} \Big[\frac{\phi(r=r_{h})}{\phi(r_{c})}\Big]^2.\label{sincleconducu}
\end{equation}
\item {\textbf{Imaginary part of conductivity $\sigma_{0} (=\Im(\lambda))$:}}  The imaginary part of the conductivity is
\begin{equation}
\Im(\lambdabf) = \frac{1}{\omega \phi^{0}}\lim_{r\rightarrow \infty}\frac {1}{2\kappa^{2}} N(r)\frac{d}{dr}\phi(r).
\end{equation}
Using eqn. (\ref{eqnmotion1}) and eqn. (\ref{eqnmotion0}), we can write
\begin{equation}
 N(r) \frac{d}{dr}\phi(r)|^{ r_{h}}_{ \infty} = -(2\kappa^{2})^{2} \rho^{2} \int^{r_{h}}_{ \infty} dr  \frac{g_{rr} g_{tt}}{\sqrt{-g} g_{xx}}\phi(r),
\end{equation}
which implies \footnote{ Let us note that, $N(r=r_{h}) = 0,$  since $N(r) = \sqrt{-g}G(r)g^{xx}g^{rr}$ and $g^{rr}(r=r_{h})=0,$ and at the boundary if $N(r\rightarrow \infty) \sim r^{1- n} $ then $\phi(r\rightarrow \infty) \sim \phi^{0}+ \phi^{1} r^{n} $.}   
\begin{equation}
\lim_{r\rightarrow \infty}N(r)\frac{d}{dr}\phi(r) = -(2\kappa^{2})^{2} \rho^{2} \int^{ r_{h}}_{\infty} dr ~~  \frac{g_{rr} g_{tt}}{\sqrt{-g} g_{xx}}\phi(r).
\end{equation}
Defining $\phi(r) = \psi(r) \phi_0$, we find
 \begin{eqnarray}
\Im(\lambdabf) &=& \frac{1}{\omega \phi^{0}}\lim_{r\rightarrow \infty}\frac {1}{2\kappa^{2}} N(r)\frac{d}{dr}\phi(r)\nonumber\\
 &=& -2\kappa^{2} \rho^{2} \int^{ r_{h}}_{\infty} dr ~~  \frac{g_{rr} g_{tt}}{\sqrt{-g} g_{xx}}\psi(r).\label{imlamda}
\end{eqnarray}
\end{itemize}

\section{Relating boundary and horizon electrical conductivity:}  In this section we consider several examples (all are asymptotically Ads spaces) and show that for each case there exist a universal relation between boundary and horizon conductivity. Rather than providing details of computation, we tabulate the results in \textbf{Table $3.1$}. For details see \cite{Son:2006em,Hartnoll:2009sz,Ge:2008ak,Jain:2009pw,Maeda:2008hn,Jain:2009uj,Hartnoll:2007ai}.
\begin{table}
 \begin{tabular}{l*{4}{c}r}
 Gravity theory in $d+1$ dimension  ~~~~~~~      & $\sigma_{H}$ ~~~~~~~~ &$\sigma_{H} (\frac{sT}{\epsilon + P})^{2}$~~~&$\sigma_{B}$ \\
\hline
 R-charge black hole in $4+1$ dim.           &$\frac{N_c^{2}T(1+k)^{2}}{16 \pi (1+\frac{k}{2})}$&$\frac{N_c^{2}T (2+k)}{32 \pi}$&$\frac{N_c^{2}T (2+k)}{32 \pi}$ \\
 R-charge black hole in $3+1$ dim.          &$\frac{N_c^{\frac{3}{2}}}{24\sqrt{2}\pi}(1+k)^{\frac{3}{2}}$&$\frac{(3+2 k)^2N_c^{\frac{3}{2}} }{6^{3}\pi\sqrt{2(1+k)}}$ &$\frac{(3+2 k)^2N_c^{\frac{3}{2}} }{6^{3}\pi\sqrt{2(1+k)}}$\\
 R-charge black hole in $6+1$ dim.          &$\frac{4 N_c^{3}T^{3}(1+k)^{3}}{81 (1+\frac{k}{3})^{3}}$&$\frac{4 N_c^{3}T^{3}(1+k)}{27(3+k)}$ &$\frac{4 N_c^{3}T^{3}(1+k)}{27(3+k)}$\\
 Reissner-Nordstrom black hole in $3+1$ dim. &$\frac{1}{g^{2}} $ &$ \frac{1}{g^{2}}(\frac{sT}{\epsilon + P})^{2}$&$\frac{1}{g^{2}}(\frac{sT}{\epsilon + P})^{2}$\\
 \end{tabular}
\caption{Real part of electrical conductivity at the horizon ($\sigma_{H}$) and at the Boundary ($\sigma_{B}$) are related by
$\sigma_{B} = \sigma_{H} \Big(\frac{s T}{\epsilon +P}\Big)^2 $. }
\end{table}
\begin{itemize}
 \item {\textbf{Single charge:}} We propose based on the observation in \textbf{Table $3.1$} that for the gauge theory with single chemical potential the expression for real part of the conductivity is given by 
\begin{eqnarray}
 \sigma_{B} &=&\frac {1}{2\kappa^{2}} G_{11}~ g_{xx}^\frac{d-3}{2}\Bigg|_{r_{h}}  \Big(\frac{s T}{\epsilon + P}\Big)^2 \nonumber\\
       &=&\sigma_{H} \Big(\frac{s T}{\epsilon +P}\Big)^2,
\label{conduc3} 
\end{eqnarray} where $s, T, P , \epsilon$ are entropy, temperature, pressure and energy density of the boundary fluid respectively. We observe that boundary conductivity can be expressed in terms of geometrical quantities evaluated at the horizon and some combination of other thermodynamic quantities.
\item {\textbf{Multiple charge:}} For multiple charge case (say there are $m$ number of chemical potential present in the gauge theory side), then boundary conductivity is $m\times m$  symmetric matrix (see \cite{Jain:2009pw}) where as horizon conductivity is $m\times m$  diagonal matrix. One can check by explicit computation that in each case
the relation
\begin{equation}\frac{1}{\rho_{I}\sigma^{-1}_{IJ}\rho_{J}} =\frac{1}{\rho_{I}\sigma^{-1}_{H, II}\rho_{I}} \Big(\frac{sT}{\epsilon +P}\Big)^{2},\end{equation} holds
 where $\sigma_{IJ}$ and  $\sigma_{H, II}$ are boundary and horizon conductivity respectively. For the action of the form eqn.(\ref{lagrangian}), the expression for horizon conductivity can be written as 
\begin{equation}
 \sigma_{H, II} =   \frac {1}{2 \kappa^{2}} G_{II}~ g_{xx}^\frac{d-3}{2}\Bigg|_{r_{h}}   .                                                                                                                                                                                         \end{equation}
 Let us note that this expression reduces to eqn.(\ref{conduc3}) in the case when single chemical potential is present.
\end{itemize}
\section{Reissner-Nordstrom black hole in arbitrary dimension:}  In this section our main focus will be on the Reissner-Nordstrom black holes in various dimensions. Besides showing the validity of eqn.(\ref{conduc3}), we will analyze the general cutoff dependence of the conductivity by choosing a hypersurface at a finite distance from the horizon. This immediately makes clear how one can prove the relation $\sigma_{B} = \sigma_{H} \Big(\frac{s T}{\epsilon +P}\Big)^2$ for more general setup. 
Action is given by 
\begin{equation}
 \int d^{d+1}x\sqrt{-g}\Big[\frac{1}{2 k^2}(R + \frac{d(d-1)}{L^2}) - \frac{1}{4 g^2}F^2\Big].
\end{equation}
The expression for the metric and gauge field for Reissner-Nordstrom black hole in arbitrary dimension are
\begin{equation}
 ds^2 =\frac{L^2}{r^2}\Big(-f(r)dt^2 + \frac{dr^2}{f(r)} + \sum\limits_{i = 1}^{d-1} dx^{i} dx^{i}\Big),\label{metricRN}
\end{equation}
and
\begin{equation}
 A_{t} = \mu \Big[1-(\frac{r}{r_{+}})^{d-2}\Big],
\end{equation} where $f(r) = 1-(1+\frac{r_{+}^{2} \mu^{2}}{\gamma^{2}}) (\frac{r}{r_{+}})^{d}+ \frac{r_{+}^{2} \mu^{2}}{\gamma^{2}} (\frac{r}{r_{+}})^{2(d-1)}$ and $\gamma^{2}=\frac{(d-1) g^{2}L^2}{(d-2)k^{2}}.$ Let us note that 
boundary is at $r = 0$ and  $\mu$ and $r_{+}$ are chemical potential and horizon radius respectively.
Various thermodynamic quantities are given by
\begin{equation}
 P = \frac{L^{d-1}}{2 k^2 r_{+}^{d}}(1+\frac{r_{+}^{2}\mu^{2}}{\gamma^{2}});~~~~~~~~~\rho = (d-1) \frac{L^{d-1}}{k^2 r_{+}^{d-2}}\frac{\mu}{\gamma^{2}}\label{Thermo},
\end{equation} and
\begin{equation}
 T=\frac{1}{4\pi r_{+}}\Big[d-\frac{(d-2)r_{+}^{2}\mu^{2}}{\gamma^{2}}\Big] ,~~~ s = \frac{2\pi}{k^2}\frac{L^{d-1}}{r_{+}^{d-1}}\label{TD2}.
\end{equation}
In order to compute the electrical conductivity we have to solve the eqn.(\ref{eqnmotion1}) for this back ground.
The eqn.(\ref{eqnmotion1}) takes the form (in $\omega \rightarrow 0$ limit)
 \begin{equation}
 \frac{d}{dr}(\frac{f(r)}{r^{d-3}}\frac{d}{dr}\phi(r)) +   \frac{2 k^2 \mu^2 (d-2)^{2} r^{d-1}}{g^2 L^{2}r_{+}^{2(d-2)}}\phi(r) = 0. 
 \end{equation}
The solution takes the form
 \begin{equation}
  \phi(r) = \phi_{0}\Bigg(1 -r^{d-2}\frac{2(d-1)(d-2)~k^2 \mu^2 r_{+}^{4-d}}{d~[g^{2}~L^{2}~(d-1)+(d-2)~k^2 \mu^2 r_{+}^{2}]}\Bigg),
 \end{equation} where $\phi_{0}$ is the boundary value of the perturbed field $\phi(r)$.
Now according to eqn.(\ref{sincleconduc})
\begin{eqnarray}
 \sigma &=& \sigma_{H}~~\Bigg(\frac{\phi(r = r_{H})}{\phi_{0}}\Bigg)^{2} \nonumber\\
      &=& \sigma_{H} \Bigg(\frac{(d-1) d g^{2}L^2 - (d-2)^{2} k^2 \mu^2 r_{+}^{2}}{d [(d-1) g^{2}L^2 + (d-2)k^2 \mu^2 r_{+}^{2}]}\Bigg)^{2}.\label{RNCON}
\end{eqnarray}
Now using the fact that $\epsilon = (d-1) P$ and the thermodynamic quantities in eqn.(\ref{Thermo}) and eqn.(\ref{TD2}) we can express 
right hand side of eqn.(\ref{RNCON}) as 
  \begin{equation}
   \sigma = \sigma_{H} \Bigg(\frac{s T}{\epsilon + P}\Bigg)^{2}\label{sigBH}.
  \end{equation}
So we have shown explicitly that for Reissner-Nordstrom black hole in any dimension, the expression for conductivity in eqn.(\ref{conduc3}) is valid.

\subsection{Cutoff dependence of conductivity:} For convenience we take the metric and gauge fields as taken in \cite{Bredberg:2010ky}.
 The asymptotically AdS black charged $p-$ brane solution are of the form
\begin{eqnarray}
 ds_{p+2}^{2} &=& -h(r)dt^{2} + \frac{dr^2}{h(r)} + e^{2t(r)}dx^{i}dx_{i},\nonumber\\
A_{t}&=& \frac{Q r_{h}}{p-1}\Bigg(1-\frac{r_{h}^{p-1}}{r^{p-1}}\Bigg),
\end{eqnarray} where 
\begin{eqnarray}
 h(r) &=& \frac{r^{2}}{R^{2}}\Bigg(1-(1 + \alpha Q^{2})\frac{r_{h}^{p+1}}{r^{p+1}} + \alpha Q^{2} \frac{r_{h}^{2 p}}{r^{2 p}}\Bigg),\nonumber\\
e^{t} &=& \frac{r}{r_{h}}.
\end{eqnarray} What we observe is that these are Reissner-Nordstrom black hole in $p+2$ dimension with gauge coupling set to one
and $\alpha = \frac{R^{2} \kappa^{2}}{p(p-1)}$. Let us consider a cutoff at radius $r = r_{c}$ outside the horizon. One can define thermodynamic quantities there. If the hawking temperature is $T_{H}$, the local temperature at the cutoff radius can expressed as 
\begin{equation}
 T_{c}\equiv T(r_{c})= \frac{T_{H}}{\sqrt{h(r_{c})}},~~~~~~ T_{H} = \frac{h^{'}(r_{h})}{4 \pi} .\label{temp}
\end{equation} The entropy density of the fluid at $r_{c}$ is given by $s = \frac{2 \pi}{\kappa^{2}} e^{-p t(r)} $, which reduces to $ s = \frac{2 \pi}{\kappa^{2}}$ as $r_{c} \rightarrow r_{h}$. 
Other thermodynamic quantities are
\begin{equation}
 \epsilon + P = \frac{\sqrt{h}}{8\pi G}\bigg(\frac{h^{'}}{2 h}-t^{'}\Bigg),\label{eprc}
\end{equation} where $\epsilon$ and $P$ are energy density and pressure of the fluid at $r_{c}.$ Let us note that for $r_{c} \rightarrow r_{h}$
\begin{equation}
 \epsilon+ P = s T_{c}.\label{TDhori}
\end{equation} The chemical potential at $r_{c}$ is~ 
\begin{equation}
 \mu = \frac{A_{t}}{\sqrt{h}},\label{chemp}
\end{equation}
~ which vanishes at the horizon. So that the thermodynamic relation
\begin{equation}
 \epsilon+ P = s T_{c} + \rho \mu,
\end{equation} holds at any arbitrary radius.
In order to find out electrical conductivity we need to solve eqn.(\ref{eqnmotion1}) for this back ground and then use eqn.(\ref{sincleconducu}) to find out conductivity at radius $r_{c}$. The solution can be obtained easily and conductivity can be written down at any radius $r_{c}$. But here we follow a slightly different route which might be helpful to generalize the results in more general background. We propose that the form of conductivity at any radius $r_{c}$ is given by
\begin{equation}
 \sigma_{c} =  \Bigg(\frac{s T}{\epsilon + P}\Bigg)^{2}\Bigg|_{r_{c}} ~\sigma_{H}~~ , \label{sigmarc}
\end{equation} where $\sigma_{c}\equiv \sigma(r_{c}), $ and $\sigma_{H}\equiv \sigma(r_{h})$. The expression for $\sigma_{H}$ is same as given in eqn. (\ref{horisig}). Let us note that,
at the boundary eqn.(\ref{sigmarc}) reproduces the desired result where as at the horizon, because of eqn.(\ref{TDhori}), $\sigma_{c}$ reduces to $\sigma_{H}$ which it should.
Comparing eqn.(\ref{sigmarc}) with eqn.(\ref{sincleconducu}), we get
\begin{eqnarray}
 \frac{\phi(r_{c})}{\phi(r_{h})} &=& \frac{\epsilon + P}{s T} \Bigg|_{r_{c}}\nonumber\\
&=& \frac{s T+ \rho \mu}{s T} \Bigg|_{r_{c}}\nonumber\\
&=& 1+\frac{\rho \mu}{s T}\Bigg|_{r_{c}},\label{solrc}
\end{eqnarray} where $\rho$ and $s$,  the charge and entropy densities are related to total charge $Q$ and entropy $S$ by a multiplicative factor of volume respectively. So we get $\frac{\rho}{s} = \frac{Q}{S}.$ It was also noted in \cite{Bredberg:2010ky}, that $S, Q$ are independent of cutoff radius $r_{c}$. Using eqn.({\ref{temp}}) and eqn.(\ref{chemp}) we get\footnote{ For the cases where $A_{t}(r_{h}) \neq 0,$ the solution takes the the form $\frac{\phi(r_{c})}{\phi(r_{h})} = 1+\frac{ \rho}{s T_{H}} [A_{t}(r_{c})-A_{t}(r_{h})]$. }
 \begin{eqnarray}
 \frac{\phi(r_{c})}{\phi(r_{h})} &=& 1+\frac{ \rho}{s T_{H}} A_{t}(r_{c})\nonumber\\
&=& 1+\frac{ Q}{S T_{H}} A_{t}(r_{c}).\label{fnsol}
 \end{eqnarray}
Now only work that is remaining is to find whether the solution of the form given in  eqn.({\ref{fnsol}}) solves eqn.(\ref{eqnmotion1}). One can very easily check that this  is indeed the case. So to summaries, the solution to eqn.(\ref{eqnmotion1}) is given by
\begin{eqnarray}
 \phi(r) &=& \frac{\epsilon + P}{s T} \Bigg|_{r} \phi(r_{h})\nonumber\\
&=&\Bigg(1+\frac{\rho}{s T} A_{t}(r)\Bigg)_{r}\Bigg(\frac{s T}{\epsilon + P}\Bigg)_{r\rightarrow \infty}\phi_{0},\label{phr}
\end{eqnarray} where $r\rightarrow \infty$ is the boundary and $\phi_{0}$ is the boundary value of $\phi.$
 The electrical conductivity for the fluid at any radius $r_{c}$ is given by
\begin{equation}
 \sigma_{c} = \Bigg(\frac{s T}{\epsilon + P}\Bigg)^{2}\Bigg|_{r_{c}} ~\sigma_{H}.
\end{equation}
\section{Imaginary part of conductivity $\sigma_{0} = \Im(\lambdabf)$:} In order to gain full knowledge of current-current correlator we need to determine the imaginary part of the electrical conductivity. As we will see, this part of the conductivity also behave in a universal way. Using eqn.(\ref{phr}) and eqn.(\ref{solrc}) we can write,
\begin{eqnarray}
 \frac{d}{dr}\phi(r) &=& \Bigg(\frac{s T}{\epsilon + P}\Bigg)_{r\rightarrow \infty}\Bigg(\frac{\rho}{s T}\Bigg)_{r\rightarrow \infty}A_{t}^{'}(r)~~ \phi_{0}\nonumber\\
&=&\Bigg(\frac{\rho}{\epsilon + P}\Bigg)_{r\rightarrow \infty}A_{t}^{'}(r)~~ \phi_{0},
\end{eqnarray} where primes denote derivative with respect to $r$.
At the boundary, imaginary part of the conductivity is given by
\begin{equation}
 \Im(\lambdabf) = \frac{1}{\omega \phi^{0}}\lim_{r\rightarrow \infty}\frac {1}{2\kappa^{2}} N(r)\frac{d}{dr}\phi(r).
\end{equation}Using eqn.(\ref{N}) and eqn.(\ref{rho}) i.e. $\rho = -\frac {1}{2\kappa^{2}}\sqrt{-g}G_{11}g^{tt}g^{rr} A_{t}^{'}(r),$
 we get
\begin{equation}
 \Im(\lambdabf) = -\frac{1}{\omega}\frac{\rho^{2}}{\epsilon + P}.\label{imlc}
\end{equation} It is interesting to compare eqn.(\ref{imlc}) with eqn.(\ref{imlamda}). Up on comparison we find,
\begin{equation}
 \frac{1}{\epsilon+ P} = 2\kappa^{2}\int^{ 1}_{0} dr ~~  \frac{g_{rr} g_{tt}}{\sqrt{-g} g_{xx}}\frac{\phi(r)}{\phi^{0}}.\label{ep}
\end{equation}
 Let us note that, in the case when $\mu = 0$, $\frac{\phi(r)}{\phi_{0}}= 1$. So we get
\begin{equation}
\frac{1}{\epsilon + P} = 2\kappa^{2}\int^{ 1}_{0} du ~~  \frac{g_{uu} g_{tt}}{\sqrt{-g} g_{xx}},
\end{equation}
 which is the result reported in \cite{Iqbal:2008by}.

Again one can study the cutoff dependence of imaginary part of the conductivity. Rather than providing details, here we write the result

\begin{equation}
 \Im(\lambdabf)_{r_{c}} =\Bigg(\frac{g_{tt}}{g_{xx}}\Bigg)_{r_{c}} \Bigg(\frac{\rho^{2}}{\epsilon + P}\Bigg)_{r\rightarrow \infty}.
\end{equation} So at the horizon, imaginary part of the conductivity vanishes (since $g_{tt}(r_{h})=0$).

\section{\textbf{Thermal conductivity:}}
Thermal conductivity is defined as the heat current response to thermal gradient in the absence of electrical current. It can be computed using the expression
\begin{equation}
   \kappa_T =\left( \frac{\epsilon+P}{ \rho }\right)^2\frac{\sigma}{T}\label{thermalconductivity1},
 \end{equation} where $\sigma $ is the electrical conductivity. For a brief discussion see appendix $A$.
Using the results of electrical conductivity discussed in earlier sections, we find

 \begin{equation}
   \kappa_T|_{r_{c}} = \sigma_{H} \frac{s^{2}}{\rho^2}T_{c},\label{KTrc}
 \end{equation} which diverges at the horizon since temperature goes to infinity at the horizon.
One interesting thing about the thermal conductivity is that $\frac{\kappa_T}{\eta T}\sum\limits_{j=1}^m(\mu^{j})^{2}$ is universal as was noticed in \cite{Jain:2009bi}. Our initial aim was to prove the universality where the the form of thermal conductivity as in eqn.(\ref{KTrc}) is very useful. But as we will see, there are other difficulties which we don't know how to tackle properly.
\begin{itemize}
 \item {\textbf{Universality of Thermal conductivity:}} In \cite{Jain:2009bi} it was observed that
\begin{eqnarray}
  \frac{\kappa_T}{\eta T}\sum\limits_{j=1}^m(\mu^{j})^{2} &=& \frac{d^\frac{\epsilon + P}{s T}2}{d-2}\Big(\frac{c^{'}}{k^{'}}\Big)\nonumber\\
   &=& 8 \pi^{2}\frac{d-1}{d^3 (d+1)}\frac{c}{k}\label{unvsalthrmalcon1}.
\end{eqnarray}
Though the proof of this statement was provided for the case of $\mu \rightarrow 0$, but there was no proof at non zero chemical potential. In what follows, we show that indeed eqn.(\ref{unvsalthrmalcon1}) is satisfied for a large class of black holes with non zero chemical potential.
\end{itemize}
 \begin{itemize}
  \item {\textbf{Proof at zero chemical potential:}} In \cite{Jain:2009bi}, a proof was provided for this case. Following \cite{Kovtun:2008kx} we can write,
\begin{equation}
  \eta  = \frac{d}{4 \pi} c^{'}T^{d-1}, ~~~~~~\sigma=\frac{1}{d-2}\frac{d}{4 \pi}k^{'} T^{d-3} \label{trnsport}.
 \end{equation}
\begin{equation}
 P = c^{'}T^{d} ,~~~~~~~~~~~~~~~~~ \chi = k^{'} T^{d-2},\label{thermo}
\end{equation} where $c^{'}$ and $k^{'}$ are related to total and charged degree of freedom of boundary gauge theory (see \cite{Kovtun:2008kx} for details).
Using eqn. (\ref{thermalconductivity1}), we can write
\begin{equation}
\frac{\kappa_T}{\eta T} \mu^{2} = \Big( \epsilon + P \Big)^{2} \frac{1}{\Big(\frac{\rho}{\mu}\Big)^{2}} \frac{1}{\Big(\frac{\eta}{\sigma T^2}\Big)}\frac{1}{T^4}.
\end{equation}
Now taking $\mu\to 0$\footnote{Let us note here that in the limit $\mu\to 0$, the charge density $\rho\to 0,$ but the ratio $\frac{\rho}{\mu}$ remains finite.  This fact makes the ratio $\frac{\kappa_T}{\eta T} \mu^{2}$ finite (non zero).}, using $\epsilon = (d-1) P $, $\chi = \frac{\rho}{\mu}$   we immediately get
 \begin{equation}
  \frac{\kappa_T}{\eta T} \mu^{2} = \frac{d^2}{d-2}\Big(\frac{c^{'}}{k^{'}}\Big).\label{unvsalthrmalcon2}
 \end{equation}
We repeat same proof using explicit metric solutions.
The metric in this case is written by
\begin{equation}
 ds^{2} = r^{2} (-f(r) dt^{2} + \sum\limits_{i=1}^{d-1} dx^{2}_{i}) + \frac{1}{f(r) r^{2}}dr^{2}.
\end{equation}

In the probe approximation we take Maxwell part of the action to be of the form,
\begin{equation}
 S = - \int d^{d+1}x~~\sqrt{-g}\frac{1}{4 g^{2}_{d+1}(r)} F_{MN}F^{MN}
\end{equation}
The charge susceptibility can be computed using (see \cite{Iqbal:2008by})
\begin{equation}
 \frac{\rho}{\mu} = \Big[\int^{\infty}_{r_h}dr~ \frac{g_{rr} g_{tt} g^{2}_{d+1}(r)}{\sqrt{-g}} \Big]^{-1}\label{romu},
\end{equation} where horizon is located at $r_h$.
Using the form of the metric and $g_{d+1}(r) = $ constant, we find
\begin{equation}
 \frac{\rho}{\mu} = \frac{d-2}{g^{2}_{d+1}} r^{d-2}_{h}.
\end{equation}
Following \cite{Iqbal:2008by} we can write, $\sigma = \frac{1}{g^{2}_{d+1}} g^{\frac{d-3}{2}}_{xx}$ and using 
\begin{equation}\epsilon + P = sT ,~~~~~~ \frac{\eta}{s} = \frac{1}{4 \pi},~~~~~~ s =  \frac{2\pi}{\kappa^{2}}g^{\frac{d-1}{2}}_{xx}\big{|}_{r=r_h},\end{equation} we reach at
\begin{eqnarray}
\frac{\kappa_T}{\eta T} \mu^{2} &=& \Big( \epsilon + P \Big)^{2} \frac{1}{\Big(\frac{\rho}{\mu}\Big)^{2}} \frac{\sigma}{\eta T^{2}}\nonumber\\
 &=& \frac{8\pi^{2} }{(d-2)^{2} }\frac{g^{2}_{d+1}}{\kappa^{2}} \frac{g^{d-2}_{xx}(r_h)}{r^{d-2}_h}\nonumber\\
 &=& \frac{8\pi^{2} }{(d-2)^{2} }\frac{g^{2}_{d+1}}{\kappa^{2}}.\label{compare}
\end{eqnarray}
 \end{itemize}
Upon comparing  eqn.(\ref{unvsalthrmalcon1}) and eqn.(\ref{compare}) gives,
\begin{equation}
 \frac{d^2}{d-2}\Big(\frac{c^{'}}{k^{'}}\Big) =  \frac{8\pi^{2} }{(d-2)^{2} }\frac{g^{2}_{d+1}}{\kappa^{2}}.
\end{equation}
\begin{itemize}
 \item \textbf{Charged Black Hole:} To prove eqn.(\ref{unvsalthrmalcon1}), we need to evaluate
\begin{equation}
 \frac{\kappa_T}{\eta T} \mu^{2} = \Big( \epsilon + P \Big)^{2} \frac{1}{\Big(\frac{\rho}{\mu}\Big)^{2}} \frac{\sigma}{\eta T^{2}}.\label{chrgebh}
\end{equation}
For general charged black holes there are few difficulties in evaluating eqn.(\ref{chrgebh}).
\begin{enumerate}
 \item { To get expression for boundary conductivity, we need to solve differential equations of the form eqn.(\ref{eqnmotion1}), which in general is not easy to solve. }
\item{For a general action (Maxwell part ) of the form
\begin{equation}
S = - \int d^{d+1}x~~\sqrt{-g}\frac{1}{4 g^{2}_{d+1}} G(r) F_{MN}F^{MN},\end{equation} the expression for the
charge susceptibility is given by \begin{equation}\frac{\rho}{\mu} = \Big[\int^{\infty}_{r_h}dr~ \frac{g_{rr} g_{tt} g^{2}_{d+1}}{G(r) \sqrt{-g}} \Big]^{-1}.
 \end{equation}Unless the explicit form of metric is known, it is not possible to compute $\frac{\rho}{\mu}$.}
\end{enumerate}

If we make use of our proposal that $\sigma = \frac{(sT)^2}{(\epsilon + P)^{2}} \sigma_{Horizon}$ , then
 \begin{eqnarray}
\frac{\kappa_T}{\eta T} \mu^{2} &=& \Big( \epsilon + P \Big)^{2} \frac{1}{\Big(\frac{\rho}{\mu}\Big)^{2}} \frac{\sigma}{\eta T^{2}}\nonumber\\ &=& \frac{4 \pi s}{(\frac{\rho}{\mu})^{2}} \sigma_{H},\label{proof}
\end{eqnarray} where we have used $\frac{\eta}{s} = \frac{1}{4 \pi} $. So we see that although we can get rid of some of the difficulties, the difficulty stated in point $2$ is not solved yet. To proceed further, we prove  eqn.(\ref{unvsalthrmalcon1}) by taking different backgrounds.
\end{itemize}

\begin{itemize}
 \item {\textbf{ Reissner Nordstrom Black Hole:}} Using the results in section $4$, we get (in the following we set $L = 1$)
 the expression for charge susceptibility,
\begin{equation}
 \frac{\rho}{\mu} = \frac{d-2}{g^{2}_{d+1}} r^{-d+2}_{h},
\end{equation}and
\begin{eqnarray}
 \sigma_{H} &=& \frac{1}{g^{2}_{d+1}} g^{\frac{d-3}{2}}_{xx}(r_h)\nonumber\\
&=& \frac{1}{g^{2}_{d+1}} \frac{1}{r_{h}^{d-3}}.
\end{eqnarray}
Now using eqn.(\ref{proof}) we get
\begin{eqnarray}
\frac{\kappa_T}{\eta T} \mu^{2} &=& \frac{8\pi^{2} }{(d-2)^{2} }\frac{g^{2}_{d+1}}{\kappa^{2}}\nonumber\\ &=& \frac{d^2}{d-2}\Big(\frac{c^{'}}{k^{'}}\Big).
\end{eqnarray}
Hence we have proved eqn.(\ref{unvsalthrmalcon1}) for arbitrary dimensional RN black hole.
\end{itemize}

\begin{itemize}
 \item {\textbf{R charged black holes in arbitrary dimension:}} In this case the Maxwell part of the action (see \cite{{Behrndt:1998jd}} for details) looks like
\begin{equation}
 S = - \int d^{d+1}x~~\sqrt{-g}\frac{1}{4 g^{2}_{d+1}} \widehat{G}(r) F_{MN}F^{MN}
\end{equation} where,
$ \widehat{G}(r) = H^{\frac{d}{d-1}}(r) $, and metric takes the form
\begin{equation}
 ds^{2} = - H^{-\frac{d-2}{d-1}}(r) f(r) dt^{2} + H^{\frac{1}{d-1}}(r) r^{2}\sum\limits_{i=1}^{d-1} dx^{2}_{i} + H^{\frac{1}{d-1}}(r)\frac{1}{f(r)} dr^{2},
\end{equation} where $H(r)$ is a harmonic function and defined by $H(r) = 1 + \frac{a}{r^{d-2}} $. Let us note that near boundary $f(r)\sim r^{2}$ and $H(r)\sim 1$. For our purpose we do not need to know the constant $a$.
Now using eqn.(\ref{romu}), we get
\begin{equation}
 (\frac{\rho}{\mu})^{-1} = g^{2}_{d+1}\int^{\infty}_{r_h}dr~ \frac{g_{rr} g_{tt} }{\widehat{G}(r) \sqrt{-g}}
\end{equation} which after some calculations yields
$\frac{\rho}{\mu} = \frac{1}{g^{2}_{d+1}}(d-2) r^{d-2}_h H(r_h)$. Using \begin{equation}\sigma_{H} = \frac{1}{g^{2}_{d+1}} \widehat{G}(r_h) g^{\frac{d-3}{2}}_{xx}(r_h),
\end{equation} and using eqn.(\ref{proof}), we get
\begin{eqnarray}
\frac{\kappa_T}{\eta T} \mu^{2} &=& \frac{8\pi^{2} }{(d-2)^{2} }\frac{g^{2}_{d+1}}{\kappa^{2}}\nonumber\\
 &=& \frac{d^2}{d-2}\Big(\frac{c^{'}}{k^{'}}\Big).
\end{eqnarray}

\end{itemize}
In the appendix $B$, we further discuss about the issue how one can avoid the difficulty as raised in point 2. 

\section{Lifshitz like black holes:} Due to possible applications in condensed matter systems, there have been lots of work \cite{Kachru:2008yh,Danielsson:2009gi,Mann:2009yx,Bertoldi:2009vn,Bertoldi:2009dt,Taylor:2008tg,others,Azeyanagi:2009pr,Li:2009pf,Brynjolfsson:2009ct,Pang:2009pd,Pang:2009wa} going on to understand transport properties of gauge theories dual to both uncharged and charged Lifshitz like black holes. Motivated by this, our aim in this section is to explore thermal and electrical conductivities for these class of black holes. This section is organized as follows. First we review the geometry and thermodynamics of uncharged Lifshitz like black holes. Then we discuss transport coefficients such as electrical conductivity. For the charged Lifshitz case, after discussing geometry and thermodynamics, we focus our attention to the computation of electrical conductivity.

\subsection{Uncharged Lifshitz black holes:}The metric for this case is given by
\begin{equation}
ds^{2}=L^{2}[-\frac{r^{2z}_{0}}{u^{2z}}f(u)dt^{2}+\frac{du^{2}}{u^{2}f(u)}
+\frac{r^{2}_{0}}{u^{2}}\sum\limits^{d}_{i=1}dx^{2}_{i}],~~~
f(u)=1-u^{z+d}.\label{metriclif}
\end{equation}
The horizon is located at $u=1$ and the boundary at
$u=0$.
We take uncharged Lifshitz black brane~(\ref{metriclif}) as the
background and treat the Maxwell action
\begin{equation}
S_{F}=-\frac{1}{4g^{2}_{d+2}}\int
d^{d+2}x\sqrt{-g}F_{\mu\nu}F^{\mu\nu}
\end{equation}
as perturbations. Here $g_{d+2}$ denotes the coupling constant. The
Maxwell equation is
\begin{equation}\frac{1}{\sqrt{-g}}\partial_{\mu}(\sqrt{-g}F^{\mu\nu})=0.\end{equation}
The electrical conductivity reads
\begin{equation}
\sigma_{B} = \frac {1}{g^{2}_{d+1}} G(u)~ g_{xx}^\frac{d-3}{2} \Big[\frac{\phi(u)}{\phi_0}\Big]^2\Big{|}_{u=1}.
\end{equation} At $\omega\rightarrow 0$ ,~~~ to get conductivity we need to solve
\begin{equation}
 \frac{d}{du}(N\frac{d}{du}\phi(u)) = 0,\label{vanmueqn}
\end{equation}
 where $N(u) $ is same as in eqn.(\ref{N}). Solution of eqn.(\ref{vanmueqn}) that is regular at the horizon is given by
$\phi(u) = \phi_{0}$, where $\phi_{0}$ is the boundary value of the perturbed field.
 Since at zero chemical potential, conductivity at the horizon and at the boundary are the same (as  $\phi(u=1)= \phi(u=0)$), we get
\begin{eqnarray}\sigma_{H} &=&\sigma_{B}\nonumber\\
&=& \frac{1}{g^{2}_{d+2}} g^{\frac{d-2}{2}}_{xx}\Big{|}_{u=1}, \end{eqnarray} which upon using metric (\ref{metriclif}), gives
\begin{equation}\sigma = \frac{1}{g^{2}_{d+2}} (L r_0)^{d-2}.\end{equation}
Using the definition of $\frac{\rho}{\mu}$ one finds,
\begin{eqnarray}\frac{\rho}{\mu} &=& \Big[\int^{\infty}_{r_0}dr~ \frac{g_{rr} g_{tt} g^{2}_{d+1}(r)}{\sqrt{-g}} \Big]^{-1}\nonumber\\&=& \frac{L^{d-2}}{g^{2}_{d+2}} (d-z) r^{d-z}_{0}.\end{eqnarray}

For these class of black hole, we have \begin{equation}\epsilon = \frac{d}{z} P , ~~~~~\epsilon + P = T s.\end{equation} Taking $P = c^{'} T^{\frac{d+z}{z}}$ we get $ s = c^{'} \frac{z+d}{z}T^{\frac{d}{z}}. $ Moreover $r^{z}_{0} = \frac{4 \pi}{z+d} T$, so that we can write
\begin{eqnarray}\chi &=&\frac{L^{d-2}}{g^{2}_{d+2}} (d-z) (\frac{4\pi}{z+d})^{\frac{d-z}{z}} T^{\frac{d-z}{z}}\nonumber\\  &=& k^{'} T^{\frac{d-z}{z}} .\end{eqnarray} In this notation the conductivity can be expressed as
\begin{equation}\sigma = k^{'}\frac{1}{d-z}(\frac{4\pi}{z+d})^{\frac{z-2}{z}} T^{\frac{d-2}{z}}.\end{equation}
We can construct interesting ratio
\begin{equation}\frac{\eta}{\sigma T^{\frac{2}{z}}} = \frac{c^{'}}{k^{'}} \frac{d-z}{z}(\frac{1}{4\pi})^{2-\frac{2}{z}} (z+d)^\frac{2(z-1)}{z}.\end{equation}
Taking $z\rightarrow 1,$ reduces to known result \cite{Kovtun:2008kx}.
Let us note that in the above we have used
\begin{equation}
c^{'}=\frac{L^{d}}{4G_{d+2}}\frac{z}{z+d} (\frac{4\pi}{z+d})^{\frac{d}{z}}, ~~~k^{'} = \frac{L^{d-2}}{g^{2}_{d+2}} (d-z) (\frac{4\pi}{z+d})^{\frac{d-z}{z}}.\label{ck}
\end{equation}

\subsection{Charged Lifshitz black holes:} It was noted in ~\cite{Taylor:2008tg} that
the following action
\begin{equation}
S=\frac{1}{16\pi G_{d+2}}\int
d^{d+2}x\sqrt{-g}(R-2\Lambda-\frac{1}{4}F^{2}-\frac{1}{2}m^{2}A^{2})
\end{equation}
admits $(d+2)-$dimensional Lifshitz space-time with arbitrary $z$
\begin{equation}
ds^{2}=L^{2}(-r^{2z}dt^{2}+\frac{1}{r^{2}}dr^{2}+r^{2}\sum\limits^{d}_{i=1}dx^{2}_{i})
\end{equation}
as a solution. If one adds a second Maxwell field ($F_{1}$) i.e.
\begin{equation}S=\frac{1}{16\pi G_{d+2}}\int
d^{d+2}x\sqrt{-g}(R-2\Lambda-\frac{1}{4}F^{2}-\frac{1}{2}m^{2}A^{2}-\frac{1}{4}F_{1}^{2}),\end{equation} then the metric of the black hole turns out to be ,
\begin{equation}
ds^{2}=L^{2}(-r^{2z}dt^{2}+\frac{1}{r^{2}}dr^{2}+r^{2}\sum\limits^{d}_{i=1}dx^{2}_{i}),~~~f(r)=1-\frac{q^{2}}{2d^{2}r^{z}}.
\end{equation}
The mass parameter and the cosmological constant
are given by
\begin{equation}
m^{2}=\frac{zd}{L^{2}},~~~\Lambda=-\frac{1}{2L^{2}}
[z^{2}+z(d-1)+d^{2}],
\end{equation}
while the massive vector field and the second Maxwell field strength
are given by
\begin{equation}
A_{t}=\sqrt{\frac{2(z-1)}{z}}
Lr^{z}f(r),~~~F_{1~rt}=qLr^{z-d-1}.
\end{equation}
Let us note that in the above $z = 2 d $ and $r_{0}^{z}\equiv q^{2}/2d^{2}$.

When $z=1$, the above ansatz leads to
\begin{equation}
ds^{2}=L^{2}[-r^{2}f(r)dt^{2}+\frac{dr^{2}}{r^{2}f(r)}+r^{2}\sum\limits^{d}_{i=1}dx^{2}_{i}],~~~
f(r)=1-\frac{m}{r^{d+1}}+\frac{q^{2}}{2d(d-1)r^{2d}},
\end{equation} and
\begin{equation}
A_{t}=0,~~~F_{rt}=0.
\end{equation}
The second Maxwell field and the cosmological constant are given by
\begin{equation}
F_{1~rt}=\frac{qL}{r^{d}},~~~\Lambda=-\frac{d(d+1)}{2L^{2}}.
\end{equation}

 In order to complete our discussion on charged Lifshitz black hole, we now discuss thermodynamics of these solutions. The temperature and
entropy are given by
\begin{equation}
T=\frac{z}{4\pi}r^{z}_{0},~~~S_{\rm
BH}=\frac{L^{d}V_{d}}{4G_{d+2}}r^{d}_{0},\label{TSZ}
\end{equation}
where $r_{0}^{z}\equiv q^{2}/2d^{2}$ and $V_{d}$ denotes the volume
of the $d-$dimensional spatial part.
Let us note that \begin{eqnarray}\chi &=& \frac{\rho}{\mu}\nonumber\\
&=& \frac{1}{16 \pi G_{d+2}}(z-d) L^{d-2}r^{d-z}_{0}\nonumber\\
&=&\frac{1}{16 \pi G_{d+2}}(z-d) L^{d-2}(\frac{4\pi}{z})^{\frac{d-z}{z}}T^{\frac{d-z}{z}},\end{eqnarray} and
\begin{equation}\rho \mu = \frac{1}{16\pi G_{d+2}}\frac{q^2 L^d}{z-d}r^{z-d}_{0}.\end{equation} Now assuming that the first law of thermodynamics is satisfied we get,
\begin{equation}
 \epsilon+ P = \frac{1}{8\pi G_{d+2}} z L^{d} r^{3d}_{0}.\label{epz}
\end{equation}
\subsection{\textbf{Electrical conductivity:}}
 In this section we shall use the coordinates $ u = (\frac{r_0}{r})^{\frac{z}{2}}$. In this coordinate,
$f(u) = 1-u^{2}$. In order to compute electrical conductivity we follow steps as discussed in section $2$ and section $3$. Using the differential eqn.(\ref{eqnmotion1}) we reach at,
\begin{equation}
 \frac{d^2}{du^2}\phi(u) + (\frac{1}{f(u)}\frac{df(u)}{du} + \frac{4-2 z}{z u})\phi(u)-\frac{2}{f(u)}\phi(u) = 0.
\end{equation}The regularized solution at the horizon takes the form
\begin{eqnarray}
&&\phi(u) = -\phi_{0}\frac{ u^{\frac{-4+3 z}{z}} \Gamma \left[-\frac{1}{2}+\frac{2}{z}\right] \Gamma \left[\frac{5}{4}-\frac{1}{z}-\frac{\sqrt{16-8 z-7 z^2}}{4 z}\right] \Gamma \left[\frac{5}{4}-\frac{1}{z}+\frac{\sqrt{16-8 z-7 z^2}}{4 z}\right]}{\Gamma \left[\frac{5}{2}-\frac{2}{z}\right] \Gamma \left[-\frac{1}{4}+\frac{1}{z}-\frac{\sqrt{16-8 z-7 z^2}}{4 z}\right] \Gamma \left[-\frac{1}{4}+\frac{1}{z}+\frac{\sqrt{16-8 z-7 z^2}}{4 z}\right]}\nonumber\\
 && \, _2F_1\left[\frac{5}{4}-\frac{1}{z}-\frac{\sqrt{16-8 z-7 z^2}}{4 z},\frac{5}{4}-\frac{1}{z}+\frac{\sqrt{16-8 z-7 z^2}}{4 z},\frac{5}{2}-\frac{2}{z},u^2\right]\nonumber\\
&&+\phi_{0} \, _2F_1\left[-\frac{1}{4}+\frac{1}{z}-\frac{\sqrt{16-8 z-7 z^2}}{4 z},-\frac{1}{4}+\frac{1}{z}+\frac{\sqrt{16-8 z-7 z^2}}{4 z},-\frac{1}{2}+\frac{2}{z},u^2\right],
\end{eqnarray}
where $\phi_{0}$ is the boundary value of $\phi(u).$ The boundary conductivity is given by
\begin{eqnarray}\sigma_{B} &=&\sigma_{H} (\frac{\phi(u=1)}{\phi(u=0)})^{2}\nonumber\\
&=& \frac{1}{16\pi G_{d+2}} (L r_0)^{d-2}(\frac{\phi(u=1)}{\phi(u=0)})^{2} .\end{eqnarray} To compute conductivity we need to calculate
$(\frac{\phi(u=1)}{\phi(u=0)})^{2}.$
\begin{itemize}
 \item {\textbf{$z=4, d=2$:}} In this case $(\frac{\phi(u=1)}{\phi(u=0)})^{2} \approx 0.24,$ so that conductivity is given by
\begin{eqnarray}
 \sigma_{B} &=&0.24~ \sigma_{H}\nonumber\\
 &=&\frac{0.24}{16\pi G_{d+2}}. \label{sigz4}
\end{eqnarray}
\item {\textbf{$z=6, d=3$:}} Here $(\frac{\phi(u=1)}{\phi(u=0)})^{2} \approx 0.27,$ which gives
\begin{eqnarray}
 \sigma_{B} &=& 0.27~ \sigma_{H}\nonumber\\ &=& \frac{0.27}{16\pi G_{d+2}} (L r_0)\nonumber\\ &=& \frac{0.27}{16\pi G_{d+2}} L (\frac{2\pi}{3})^{\frac{1}{6}}T^{\frac{1}{6}}.\label{sigz6}
\end{eqnarray}
\end{itemize}
In general the conductivity can be written as
\begin{equation}
 \sigma_{B} = \frac{C}{16\pi G_{d+2}} L^{d-2} (\frac{4\pi}{z})^{\frac{d-2}{z}}T^{\frac{d-2}{z}},\label{genlifcon}
\end{equation}
where $C = (\frac{\phi(u=1)}{\phi(u=0)})^{2}.$ It is important to note that above expressions only depends on temperature (no dependence in chemical potential), since charge and temperature are related by
$ T = \frac{q^{2}}{2\pi z}$.

Let us note that, using eqn.(\ref{TSZ}) and eqn.(\ref{epz}), eqn.(\ref{sigBH}) gives
\begin{eqnarray}
 \sigma_{Boundary} &=& \frac{(sT)^2}{(\epsilon + P)^{2}} \sigma_{Horizon}\nonumber\\
 &=& 0.25~~ \sigma_{H},\label{HBCZ}
\end{eqnarray} which is independent of $z, d .$ What we observe is that, electrical conductivity of charged Lifshitz like black holes given in eqn.(\ref{sigz4}) and in eqn.(\ref{sigz6}) differs slightly from eqn.(\ref{HBCZ}).

\subsection{Thermal conductivity for Lifshitz like theories:} In the following we discuss thermal conductivity and thermal conductivity to viscosity ratio for both uncharged and charged Lifshitz like black holes.
\begin{itemize}
 \item \textbf{{Uncharged Lifshitz black holes:}} Now we can construct the ratio $\frac{\kappa_{T} \mu^{2}}{\eta T^{3-\frac{2}{z}}} $, which after computation gives
\begin{equation}
\frac{\kappa_{T} \mu^{2}}{\eta T^{3-\frac{2}{z}}} = (4 \pi)^{2-\frac{2}{z}}(z+d)^{\frac{2}{z}}\frac{1}{z (d-z)}\frac{c^{'}}{k^{'}} .\label{lifunkt}
\end{equation}
By taking $z = 1$, above ratio reduces to \begin{equation}\frac{\kappa_{T} \mu^{2}}{\eta T} = \frac{(d+1)^{2}}{d-1}\frac{c^{'}}{k^{'}}. \end{equation} To match with the result of the previous section one simply has to make the substitution $d\rightarrow (d-1)$ in order to match the convention of previous sections (compare metrics).
\item \textbf{{Charged Lifshitz black holes:}}The  thermal conductivity can be computed using
\begin{equation}\kappa_{T} = \left( \frac{\epsilon+P}{ \rho }\right)^2\frac{\sigma}{T}.\end{equation}
For charged background we have, \begin{equation}P = c^{'} T^{\frac{d+z}{z}}f_{p}(\frac{\mu}{T}),~~~~ \chi = k^{'}T^{\frac{d-z}{z}}f_{\chi}(\frac{\mu}{T}),\end{equation} where
\begin{equation}c^{'}=\frac{L^{d}}{4G_{d+2}}\frac{z}{z+d} (\frac{4\pi}{z+d})^{\frac{d}{z}}, ~~~k^{'} = \frac{L^{d-2}}{16 \pi G_{d+1}} (d-z) (\frac{4\pi}{z+d})^{\frac{d-z}{z}}.\end{equation}
Using all the above formulas and eqn.(\ref{genlifcon}) we get,
\begin{equation}
\frac{\kappa_{T} \mu^{2}}{\eta T^{3-\frac{2}{z}}} =  2^{\frac{2}{z}} 3^{2-\frac{2}{z}} C (4 \pi)^{2-\frac{2}{z}}(z+d)^{\frac{2}{z}}\frac{1}{z (z-d)}\frac{c^{'}}{k^{'}}.\label{lifkt}
\end{equation}
 For the case $z=6, d=3$ , we have $C = 2.12$ and for $z=4, d=2$ we have $C = 1.76$.
  Let us note that, if we compare eqn.(\ref{lifkt}) with eqn.(\ref{lifunkt}) we see that charged and uncharged cases are same up to some numerical factor. Let us also note that for the case of asymptotically AdS charged or uncharged black holes, the thermal conductivity to viscosity ratio were same.
\end{itemize}

\section{Conclusion:} At zero chemical potential, it was shown in \cite{Iqbal:2008by}, that expression for electrical conductivity\footnote{Let us note that,  at zero chemical potential, authors in  \cite{Hassanain:2010fv},  deduced the full set of five-dimensional operators induced by
stringy corrections which are  
present 
in any theory that has an $AdS_5 * M_5$ structure, and computed the electrical conductivity.} is given by
  \begin{equation}\sigma = \frac {1}{g^{2}_{d+1}}  g_{xx}^\frac{d-3}{2}\Big{|}_{r=r_h},\label{sigend}\end{equation} given that the form of 
 Maxwell part of the action is
\begin{equation} S = - \int d^{d+1}x~~\sqrt{-g}\frac{1}{4 g^{2}_{d+1}}  F_{MN}F^{MN}\label{max1}.\end{equation}  It was also argued  that radial evolution of electrical conductivity at zero chemical potential is trivial, as a result evaluating conductivity at any radial position gives same result. As we have seen, at non-zero chemical potential ($\mu$) , radial evolution is nontrivial.  Based on few observations, we have proposed that, at $\mu \neq 0$, given that the form of 
 Maxwell part of the action is
\begin{equation} S = - \int d^{d+1}x~~\sqrt{-g}\frac{1}{4 g^{2}_{d+1}} \widehat{G}(r) F_{MN}F^{MN}\label{max2},\end{equation} the  electrical conductivity at the boundary is given by 
\begin{eqnarray}\sigma_{B} &=& \frac {1}{g^{2}_{d+1}} \widehat{G}(r) g_{xx}^\frac{d-3}{2}\Big{|}_{r=r_h} \frac{(sT)^{2}}{(\epsilon+ P)^{2}}\nonumber\\
&=& \sigma_{H}\frac{(sT)^{2}}{(\epsilon+ P)^{2}}\label{HBC},
\end{eqnarray} where $\sigma_{H} = \frac {1}{g^{2}_{d+1}} \widehat{G}(r) g_{xx}^\frac{d-3}{2}\Big{|}_{r=r_h}$, is the electrical conductivity radially evaluated at the horizon\footnote{Let us note that, at $\mu =0$, (and taking $\widehat{G}(r) =1 $ to match the form of eqn.(\ref{max2}) with eqn.(\ref{max1})) eqn.(\ref{HBC}) reduces to  eqn.(\ref{sigend})}.
 We have also  computed the cutoff dependent conductivity which interpolates smoothly between the results at the horizon and at the boundary. We have also shown that, once the real part of the conductivity is known, the imaginary part of conductivity is automatically fixed. To summaries, in the presence of chemical potential the electrical conductivity \footnote{Let us note that for extremal black holes also, electrical conductivity shows some universality, see \cite{Edalati:2009bi},\cite{Jain:2009pw}.} can be expressed as
\begin{equation}
 \lambdabf = -\frac{1}{\omega}\frac{\rho^{2}}{\epsilon + P} + \frac {1}{g^{2}_{d+1}} \widehat{G}(r)~ g_{xx}^\frac{d-3}{2}\Big{|}_{r=r_h} \frac{(sT)^{2}}{(\epsilon+ P)^{2}}.
\end{equation}

  Using eqn.(\ref{HBC}), we could show that, for sufficiently general background 
\begin{equation}\frac{\kappa_T}{\eta T}\sum\limits_{j=1}^m(\mu^{j})^{2} = \frac{d^2}{d-2}\Big(\frac{c^{'}}{k^{'}}\Big).\end{equation} We also computed the electrical conductivity and then thermal conductivity to viscosity ratio for  both uncharged and charged Lifshitz like black holes. In the case of charged Lifshitz like black holes, eqn.(\ref{HBC}) does  not hold . The value of thermal conductivity to viscosity ratio differs from that of uncharged case by some numerical factor. Let us note that,
\begin{equation}
 P = c^{'}T^{\frac{d+z}{z}} f_{P}(\frac{\mu}{T}) ,~~~~~~~~~~~~~~~~~ \chi = k^{'} T^{\frac{d-z}{z}}f_{\chi}(\frac{\mu}{T}).\label{thermocon}
\end{equation} Now for charged Lifshitz black holes, since chemical potential ($\mu$) and temperature $T$ are related , functions $f_{p}, f_{\chi}$'s becomes numbers. This implies finding out $c^{'}$ and $k^{'}$ are problematic. If we account for these facts carefully, thermal conductivity to viscosity ratio for charged and uncharged case might agree.

\section*{Acknowledgments}
  It is pleasure to thank Sudipta Mukherji for various discussions and taking personal care in order to improve the presentation. I am thankful to Sean Hartnoll for giving excellent set of lectures at the Mahabaleswar school, for various helpful comments on the earlier version of the manuscript and for encouragement.I am also thankful to  Balram Rai for useful discussions and other members of string group at IOP for encouragement. 
\renewcommand{\thesection}{\Alph{section}}
\setcounter{section}{0}
\section{Relation between universal conductivity of stretched horizon and boundary conductivity:} 
 Consider the Maxwell part of the action of the form
\begin{equation} S = - \int d^{d+1}x~~\sqrt{-g}\frac{1}{4 g^{2}_{d+1}(r)}  F_{MN}F^{MN},\end{equation} where $g^{2}_{d+1}(r)$ in general is a $r$ dependent coupling\footnote{Let us note that, in the notation used in eqn.(\ref{lagrangian}), $\frac{1}{4 g^{2}_{d+1}(r)}\equiv \frac{1}{4 g_{d+1}^{2}} \widehat{G}(r)$}. The electrical conductivity at any radius is given by (see eqn. (\ref{sincleconducu}) and for further details see \cite{Myers:2009ij,Jain:2009pw})
\begin{equation}
 \sigma(r_{c}) =\Bigg(\frac{1}{4 g^{2}_{d+1}(r)}~ g_{xx}^\frac{d-3}{2}\Bigg)_{r=r_{h}} \Big[\frac{\phi(r=r_{h})}{\phi(r_{c})}\Big]^2.\label{cr}
\end{equation}
Let us note that at the horizon conductivity is
\begin{equation}
 \sigma(r_{c}=r_{h}) =\Bigg(\frac{1}{4 g^{2}_{d+1}(r)}~ g_{xx}^\frac{d-3}{2}\Bigg)_{r=r_{h}},\label{ch}
\end{equation} which is 
 entirely given by geometrical quantities evaluated at the horizon. In order to understand radial evolution of conductivity we consider the cases with vanishing and non-vanishing chemical potential separately. 
\subsection{Radial evolution of conductivity at zero chemical potential} Let us note that at vanishing chemical potential, the term $M(r) = 0$ in  eqn. (\ref{eqnmotion1}). If we impose ingoing boundary condition at the horizon and impose $\phi(r\rightarrow \infty) = \phi_{0}$, at the boundary , then one can show that solution to  eqn. (\ref{eqnmotion1}) is given by $\phi(r) = \phi_{0}$, at any radius i.e. $\phi$ is a constant. Now using eqn.(\ref{cr}) we get,
\begin{eqnarray}
 \sigma_{\mu = 0}(r_{c}) &=&\Bigg(\frac{1}{4 g^{2}_{d+1}(r)}~ g_{xx}^\frac{d-3}{2}\Bigg)_{r=r_{h}} \nonumber\\
&=& \sigma_{\mu = 0}(r_{h}). 
\end{eqnarray} So we conclude that at vanishing chemical potential boundary and horizon conductivity is the same.
\begin{itemize}
 \item \textbf{Relation with universal conductivity of the stretched horizon:} The universal conductivity of the stretched horizon is given by (see \cite{Iqbal:2008by})
$\sigma_{mb} =\frac{1}{g^{2}_{d+1}(r_{h})}  $. Now we see
\begin{equation}
 \sigma_{CFT, \mu =0} = \sigma_{mb}~g_{xx}^{\frac{d-3}{2}}(r_{h}),\label{CFTmb}
\end{equation} where factor $g_{xx}^{\frac{d-3}{2}}(r_{h})$ converts the length scale in CFT to proper length scale at horizon \cite{Iqbal:2008by,Myers:2009ij } (let us note that in $d$ dimension, conductivity has a mass dimension $d-3$).\end{itemize}
\subsection{Radial evolution of conductivity at finite chemical potential} At non zero chemical potential we have (assuming $A_{t}(r_{h}) = 0$ so that $A_{t}(r\rightarrow \infty) = \mu $ , see eqn.(\ref{fnsol}))
\begin{equation}
 \frac{\phi(r_{c})}{\phi(r_{h})} = 1+\frac{ \rho}{s T_{H}} A_{t}(r_{c}).
\end{equation}
 Using eqn.(\ref{cr}) we get,
\begin{equation}
 \sigma_{\mu \neq 0}(r_{c}) =\Bigg(\frac{1}{4 g^{2}_{d+1}(r)}~ g_{xx}^\frac{d-3}{2}\Bigg)_{r=r_{h}} \Bigg(\frac{s T_{H}}{s T_{H} + \rho A_{t}(r_{c}) }\Bigg)^{2},\label{rcm}
\end{equation} which implies
\begin{equation}
 \sigma_{\mu \neq 0}(r_{h}) =\Bigg(\frac{1}{4 g^{2}_{d+1}(r)}~ g_{xx}^\frac{d-3}{2}\Bigg)_{r=r_{h}} ,\label{rcmh}
 \end{equation} and 
\begin{eqnarray}
 \sigma_{\mu \neq 0}(r\rightarrow \infty)& =&\Bigg(\frac{1}{4 g^{2}_{d+1}(r)}~ g_{xx}^\frac{d-3}{2}\Bigg)_{r=r_{h}}\Bigg(\frac{s T_{H}}{\epsilon + P}\Bigg)^{2}\nonumber\\
&=&  \sigma_{\mu \neq 0}(r_{h}) \Bigg(\frac{s T_{H}}{\epsilon + P}\Bigg)^{2}. \label{rcmb}
 \end{eqnarray}
So we conclude that at finite chemical potential boundary and horizon conductivity are different and there is a simple relation between them.
\begin{itemize}
 \item \textbf{Relation with universal conductivity of the stretched horizon at finite chemical potential:} Again the universal conductivity of the stretched horizon is given by 
$\sigma_{mb} =\frac{1}{g^{2}_{d+1}(r_{h})}  $. Once again we observe that,
\begin{equation}
 \sigma_{\mu \neq0}(r_{h}) = \sigma_{mb}~g_{xx}^{\frac{d-3}{2}}(r_{h}),
\end{equation} and 
\begin{equation}
 \sigma_{CFT,\mu \neq0}(r\rightarrow \infty) = \sigma_{mb}~g_{xx}^{\frac{d-3}{2}}(r_{h})\Bigg(\frac{s T_{H}}{\epsilon + P}\Bigg)^{2}.\label{CFTmbmu}
\end{equation} Let us note that at $\mu = 0$, eqn.(\ref{CFTmbmu}) reduces to  eqn.(\ref{CFTmb}).
\end{itemize}

\section{Thermal conductivity:}
In the following we first review the hydrodynamics with multiple conserved charges (see \cite{Jain:2009pw}) and write down an expression for thermal conductivity. The single charge case was discussed in  \cite{Son:2006em} .

\begin{itemize}
 \item \textbf{\textit{Relativistic hydrodynamics with multiple conserved charge:}}
 The continuity equations are normally presented as
\begin{equation}
   \d_\mu T^{\mu\nu}=0, ~~~ \d_\mu J_{i}^\mu =0
\end{equation}
where
\begin{equation}
   T^{\mu\nu} = (\epsilon+P) u^\mu u^\nu + P g^{\mu\nu}+\tau^{\mu\nu}
    ,~~~
    J_{i}^\mu = \rho_{i} u^\mu + \nu_{i}^\mu\label{velo}
\end{equation}
In the above $\epsilon$ and P are the local energy density and pressure respectively, $u^\mu$ is the local velocity and it obeys $ u_\mu u^\mu = -1$, where as $\tau^{\mu\nu}$ and $\nu_{i}^\mu$ are the dissipative parts of stress-energy tensor and current.

 One can choose
 $u^\mu$ and $\rho_{i}$'s so that
\begin{equation}\label{transverse}
    u_\mu \tau^{\mu\nu} = u_\mu \nu^\mu_{i} = 0\,.
\end{equation}

We also have
\begin{equation}
    \epsilon+P = Ts +\sum\limits_{i=1}^m \mu^{i} \rho_{i}, \qquad d\epsilon = Tds +\sum\limits_{i=1}^m \mu^{i} d\rho_{i}.\label{thermo1}
\end{equation}
One can write
\begin{equation}
 \nu^{\mu}_i = -\sum\limits_{j=1}^m\varkappa_{ij} \left(\d^\mu\frac{\mu^{j}}{ T}+ u^\mu u^\lambda \d_\lambda \frac{\mu^{j}}{ T}\right)
\end{equation} and similarly for $\tau^{\mu\nu}$ (see \cite{Son:2006em}) .
 To interpret $ \varkappa_{ij} $  as the
 coefficient of thermal conductivity, consider no charge current i.e. $ J^\alpha_{j} = 0, $\footnote{In our notation $\mu ,\nu$ runs from $t,1,2...D$ , where as $\alpha$ runs from $1,2,...D$, and i,j are R-charge indices.}  but there is an energy flow,
  $ T^{t\alpha}\neq 0 $, which is the heat flow.  Take $ u^\alpha $ to be small so that one gets using eqn. (\ref{velo})
\begin{equation}
 \rho_{i} u^\alpha=\sum\limits_{j=1}^m\varkappa_{ij} \d^\alpha\frac{\mu^{j}}{T}.
\end{equation}

From which one can reach at

  \begin{eqnarray}
     T^{t\alpha} &=& (\epsilon+P) u^\alpha\nonumber\\
     &=& -\frac{1}{\sum\limits_{i,j=1}^m\rho_{i}\varkappa_{ij}^{-1}\rho_{j}}(\frac{\epsilon+P}{T})^{2}(\d^\alpha T-\frac{T}{\epsilon+P}\d^\alpha P),
 \end{eqnarray}
hence the coefficient of thermal conductivity can be read off as
 \begin{equation}
   \kappa_T =\left( \frac{\epsilon+P}{ T}\right)^2\frac{1}{\sum\limits_{i,j=1}^m\rho_{i}\varkappa_{ij}^{-1}\rho_{j}}\label{thermalconductivity}.
 \end{equation}
Note that, $\varkappa_{ij}$ can be found out from greens function as
\begin{eqnarray}
   G_{ij}^{xx} (\omega, q=0) &=& -i \omega \frac{\varkappa_{ij}}{ T}\ \nonumber\\ &=& -i \omega \sigma_{ij},
 \end{eqnarray}
where $ J^{x}_{i}=-G_{ij}(\omega,q=0)A^{x}_j $  and $ \sigma_{ij}(\omega,q=0)$ can be obtained using current-current correlator as discussed earlier.

\end{itemize}

Let us note that for single charge black hole $\frac{1}{\rho_{i}\varkappa_{ij}^{-1}\rho_{j}}=\frac{\varkappa}{\rho^{2}}$. Therefore
\begin{eqnarray}
   \kappa_T &=&\left( \frac{\epsilon+P}{ \rho T}\right)^2\varkappa \nonumber\\&=&\left( \frac{\epsilon+P}{ \rho }\right)^2\frac{\sigma}{T}.\label{TCD}
 \end{eqnarray}

\section{Proving universality of thermal conductivity:}
As was discussed in the text that there are few problems in order to prove the thermal conductivity to viscosity ratio. Here we discuss how one can over come that difficulty.  
\begin{itemize}
 \item {\textbf{General single charged black hole:}} The metric that we want to work with is of the form
\begin{equation}
 ds^{2} = - e^{-2(d-2) U(r)} f(r) dt^{2} +   e^{2 U(r)} ( r^{2} d\Omega^{2}_{d-1} +\frac{dr^{2}}{f(r)} ).
\end{equation}
The Maxwell part of the action takes the form
\begin{equation}
 S = - \int d^{d+1}x~~\sqrt{-g}\frac{1}{4 g^{2}_{d+1}} G(r) F_{MN}F^{MN}.
\end{equation}
 Using eqn. (\ref{romu}) we get,
\begin{eqnarray}
 (\frac{\rho}{\mu})^{-1} &=& g^{2}_{d+1}\int^{\infty}_{r_0}dr~ \frac{g_{rr} g_{tt} }{G(r) \sqrt{-g}}\nonumber\\ &=& g^{2}_{d+1}\int^{\infty}_{r_0}dr~\frac{e^{2(2-d) U(r)}}{G(r) r^{d-1}}.\label{chigen}
\end{eqnarray} Let us note that, unless we know explicitly the functions $G(r), U(r)$, it is not possible to evaluate R.H.S. of eqn.(\ref{chigen}). However, here we can make use of first order formalism of supergravity, which we are going to review below (see \cite{Cardoso:2008gm} and references there in for details). For simplicity we concentrate on five dimension i.e. $d=4$. The relevant part of the five-dimensional action is
\begin{equation}
S = \frac{1}{16 \pi \, G_5} \int d^5x \sqrt{-g} \left( R
-  g_{ij} \, \partial_M \varphi^i \partial^M \varphi^j
- \frac12 G_{AB} \,F^A_{MN} F^{B \,MN } -
V_{\rm pot}
\right)
\;.
\label{bulkaction5}
\end{equation}
Space-time metric is $g_{MN}$.
The real scalar fields $X^A$ satisfy the constraint
\begin{equation}
\frac16 \, C_{ABC} \, X^A \, X^B \, X^C =1 \;.
\label{vsgconstraint}
\end{equation}
The metric $G_{AB}$ is given by
\begin{equation}
G_{AB} = - \frac12 \, C_{ABC} \, X^C + \frac92 \, X_A \, X_B \;,
\end{equation}
where
\begin{equation}
X_A = \frac16 \, C_{ABC} \, X^B \, X^C \;.
\end{equation}
Let us note that $X^A \, X_A =1,$ which follows from \eqref{vsgconstraint}.

Following \cite{Behrndt:1998jd} we consider non-extremal electrically charged
static black hole solutions
\begin{equation}
ds^2_5 = - {\rm e}^{-4 U} \, f \, dt^2 + {\rm e}^{2 U} \, f^{-1} \, dr^2
+  {\rm e}^{2 U} \, r^2 \, d \Sigma_k^2 \;\;\;,\;\;\;
f = k - \frac{m}{r^2} + \mathfrak{g}^2 \, r^2 \, {\rm e}^{6 U} \;,
\label{bhline}
\end{equation}
where $U=U(r), f=f(r)$.
Here $d \Sigma_k^2$ denotes the line element of
a three-dimensional space of constant curvature with metric
$\eta_{\alpha \beta}$, either flat space ($k=0$), hyperbolic space ($k=-1$)
or a unit three-sphere $S^3$ ($k=1$).
The presence of a non-vanishing parameter $m$
is necessary in order for the solutions to have a horizon.
We also have
\begin{equation}
X_A = \frac{1}{3} \, e^{-2 U} \, H_A \;,
\label{xa}
\end{equation}as well as
\begin{equation}
F^A_{tr} = e^{-4U} \, \frac{G^{AC}  Q_C}{r^3} \;,
\label{fieldstr}
\end{equation} where $H_A$ is given by
\begin{equation}
H_A = h_A + \frac{q_A}{r^2} \;.
\label{harmf}
\end{equation}
 The physical electric charges $Q_A$
are related to the $q_A$ by
\begin{equation}
Q_A G^{AB} Q_B = k \, q_A G^{AB} q_B + m \, q_A G^{AB} h_B \;.
\label{Qq}
\end{equation}

If we write $F_{tr}^A = - \partial_r {\phi}^A (r)$, then
\begin{equation}
Q_A \, {\phi}^A =  - \left(k - \frac{m}{r^2}\right) \, e^{-2U} \,
q_A X^A + k \, q_A h^A \;.
\label{qmu}
\end{equation}We chose the integration constant in such a way that $Q_A \,
{\phi}^A$ vanishes
at spatial infinity.

\end{itemize}
\begin{itemize}
 \item \textbf{Single charge:} Let us take planer black hole i.e. $k=0$. Let us also note that, in our notation physical charge density is $\rho_{A} = \frac{1}{4\pi G} Q_{A}.$
Using eqn. (\ref{qmu}) and eqn. (\ref{Qq}) we can write,
\begin{eqnarray}
 \frac{\mu}{\rho} &=&\frac{\phi}{\rho}\Big{|}_{r_{h}}\nonumber\\&=& 4 \pi G_{5} e^{-2U(r_h)} X^1 \frac{1}{r^{2}_{h}}\Big{|}_{r_{h}}.\label{phiro}
\end{eqnarray}
The expression for horizon conductivity is $\sigma_{H} = \frac{1}{4\pi G}G_{11} g^{\frac{1}{2}}_{xx}\Big{|}_{r=r_h}.$
The thermal conductivity to viscosity ratio reads,
\begin{eqnarray}
 \frac{\kappa_T}{\eta T} \mu^{2}  &=& \frac{4 \pi s}{(\frac{\rho}{\mu})^{2}} \sigma_{H}\nonumber\\
 &=& \frac{1}{4 G^{2}_{d+1}}G_{11} g^{2}_{xx}(\frac{\mu}{\rho})^{2}\Big{|}_{r=r_h}.\label{TC}
\end{eqnarray}
Using $h=1$, $G_{11} =\frac{1}{2 (X^{1})^{2}}$  and eqn. (\ref{phiro}), eqn. (\ref{TC}) gives
 \begin{eqnarray}\frac{\kappa_T}{\eta T} \mu^{2}&=& 4 \pi^{2}G_{11} (X^{1})^{2}\Big{|}_{r=r_h}\nonumber\\ &=& 2 \pi^{2}.\end{eqnarray}
\end{itemize}
\begin{itemize}
 \item \textbf{Multi charge:} Here we take $G_{AB}$ to be diagonal for simplicity. Note that in this case the expression for boundary conductivity is a matrix and contains off diagonal terms but horizon conductivity is diagonal and proportional to $G_{II}$.  We propose the following to hold.
\begin{equation}\frac{1}{\rho_{I}\sigma^{-1}_{IJ}\rho_{J}} =\frac{1}{\rho_{I}\sigma^{-1}_{H, II}\rho_{I}} (\frac{sT}{\epsilon +P})^{2}.\end{equation}
Since \begin{equation}\sigma_{H,II} = \frac{1}{4\pi G_{d+1}} G_{II}~ g_{xx}^\frac{d-3}{2}\Big{|}_{r=r_h}\end{equation} and $G^{-1}_{II} = G^{II}$ we get,
\begin{equation}\rho_{I}\sigma^{-1}_{H, II}\rho_{I} = 4\pi G_{d+1}~ g_{xx}^\frac{3-d}{2}\rho_{I}G^{II}\rho_{I} \Big{|}_{r=r_h}.\end{equation} As $G_{AB}$ is diagonal, we can also write
$\mu q_{I} = Q^{2}_{I} $ so that \begin{eqnarray}\frac{\phi^{I}}{\rho^{I}} &=& 4 \pi G \frac{\mu}{r^{2}_{+}}e^{-2U(r_{h})}\frac{q_{I} X^{I}}{Q^{2}_{I}}\nonumber\\ &=& 4 \pi G \frac{X^{I}}{r^{2}_{+}}e^{-2U(r_{h})}.\end{eqnarray} As a result we obtain,
\begin{equation}\sum\limits_{I=1}^m (\phi^{I})^{2} = \frac{1}{r^{4}_{+}}e^{-4 U(r_h)}\sum\limits_{I=1}^m Q^{2}_{I} (X^{I})^{2}.\end{equation} Using $G^{II} = 2 (X^{I})^{2}$, we get
\begin{equation}
 \sum\limits_{I=1}^m (\phi^{I})^{2} =\frac{1}{2 r^{4}_{h}}e^{-4 U(r_h)}\sum\limits_{I=1}^m Q^{2}_{I} G^{II}.\label{summu}
\end{equation}

The thermal conductivity to viscosity ratio reads,
\begin{eqnarray} \frac{\kappa_T}{\eta T}\sum\limits_{I=1}^m(\phi^{I})^{2} &=& \frac{(\epsilon + P)^{2}}{\eta T^{2}}\frac{\sum\limits_{I=1}^m(\phi^{I})^{2}}{(\frac{1}{4\pi G})^{2}Q_{I}\sigma^{-1}_{IJ} Q_{J}}\nonumber\\ &=& 4\pi s g^{\frac{1}{2}}_{xx} \frac{\sum\limits_{I=1}^m(\phi^{I})^{2}}{\frac{1}{4\pi G_{d+1}}Q_{I}G^{IJ} Q_{J}}\Big{|}_{r=r_h} . \end{eqnarray} Now using eqn. (\ref{summu})  and  $ s= \frac{1}{4 G_{d+1}} g^{\frac{3}{2}}_{xx}\Big{|}_{r=r_h}$ we reach at,
\begin{eqnarray} \frac{\kappa_T}{\eta T}\sum\limits_{I=1}^m(\phi^{I})^{2} &=& 4\pi^{2} g^{2}_{xx}(r_h)\frac{1}{2 r^{4}_{h}}e^{-4 U(r_h)}\nonumber\\ &=& 2\pi^{2} .\end{eqnarray}
\end{itemize}
Let us note that,\begin{eqnarray}\frac{d^2}{d-2}\Big(\frac{c^{'}}{k^{'}}\Big) &=& \frac{\pi }{(d-2)^{2} }\frac{g^{2}_{d+1}}{G_{d+1}}\nonumber\\ &=& 2 \pi^{2},\end{eqnarray} since
$g^{2}_{d+1} =8\pi G $ and $d=4$. Generalizing the above results for the cases with arbitrary $G_{AB}$ is interesting but not straight forward.

\section{Thermal conductivity for non abelian black hole:} In this section we discuss thermal conductivity to viscosity ratio for black hole with non abelian gauge fields. The gravity background is written in \cite{Torabian:2009qk}.
The action contains
gravity with $SU(2)$ gauge fields\footnote{Let us note that $\epsilon^{abc}$
below can be changed to any other structure constants of a Lie algebra to have the general results for arbitrary Lie groups.}
\begin{equation}
{\cal L}=\frac{1}{16\pi G_{d+1}}\left( R+d(d-1) -F^a_{MN}F^{aMN}\right)\quad,
\end{equation}
with
\begin{equation}
F_{MN}^a = \partial_M A^a_N- \partial_N A^a_M+\epsilon^{abc}A^b_M A^c_N\quad.
\end{equation}
A charged black-brane solution in a general boosted frame is
\begin{equation}
ds^2= -r^2 V(r)u_\mu u_\nu dx^\mu dx^\nu -2 u_\mu dx^\mu dr + r^2\left(\eta_{\mu\nu}+u_\mu u_\nu\right)
dx^\mu dx^\nu\quad,
\end{equation}
\begin{equation}
A^a= \sqrt{\frac{d-1}{2(d-2)}}\frac{q^a}{r^{d-2}} u_\mu dx^\mu\quad,
\end{equation}
with
\begin{equation}
V(r)=1-\frac{m}{r^n}+\frac{q^a q^a}{r^{2d-2}}\quad.
\end{equation}
where it is simply obtained by embedding the $U(1)$ Reissner-Nordstrom black-brane into
a Cartan direction inside $SU(2)$ which is specified by $q^a$ ($a=1,2,3$).
The conserved $SU(2)$ currents are
\begin{eqnarray}
J^{\mu a(0)} &=& \frac{1}{4\pi G_{d+1}}\sqrt{\frac{(d-1)(d-2)}{2}}q^a u^\mu \nonumber\\ &\equiv& \rho^a u^\mu \label{0thdensity}
  \quad.
\end{eqnarray}
So charge density and chemical potential reads,
 \begin{eqnarray}\rho^a &=& \frac{1}{4\pi G_{d+1}}\sqrt{\frac{(d-1)(d-2)}{2}}q^a,~~\mu^{a} \nonumber\\&=& \sqrt{\frac{(d-1)}{2(d-2)}}\frac{q^a}{r^{d-2}_{h}}.\end{eqnarray}
Since
\begin{equation}\frac{1}{\rho_{a}\sigma^{-1}_{ab}\rho_{b}} =\frac{1}{\rho_{a}\sigma^{-1}_{H, aa}\rho_{a}} (\frac{sT}{\epsilon +P})^{2}\end{equation} and
 $\sigma_{H,aa} = \frac{1}{4\pi G_{d+1}} ~ g_{xx}^\frac{3-d}{2}\Big{|}_{r=r_h}$, we get
 \begin{eqnarray} \frac{\kappa_T}{\eta T}\sum\limits_{a=1}^3(\mu^{a})^{2} &=& \frac{1}{4 G^{2}_{d+1}}\frac{\sum\limits_{a=1}^3(\mu^{a})^{2}}{\sum\limits_{a=1}^3(\rho^{a})^{2}}\nonumber\\ &=& \frac{4 \pi^{2}}{(d-2)^{2}}.\end{eqnarray}
Let us note that,\begin{eqnarray}\frac{d^2}{d-2}\Big(\frac{c^{'}}{k^{'}}\Big) &=& \frac{\pi }{(d-2)^{2} }\frac{g^{2}_{d+1}}{G_{d+1}}\nonumber\\ &=& \frac{4 \pi^{2}}{(d-2)^{2}},\end{eqnarray} since
$g^{2}_{d+1} =4\pi G_{d+1}. $

\end{document}